\documentstyle[prb,twocolumn,floats,epsf,aps]{revtex}
\begin{document}
\draft

\twocolumn[\hsize\textwidth\columnwidth\hsize\csname @twocolumnfalse\endcsname

\title{Stochastic series expansion method with operator-loop update}

\author{Anders W. Sandvik}
\address{Department of Physics, University of Illinois at Urbana-Champaign,
1110 West Green Street, Urbana, Illinois 61801 \\
and Center for Nonlinear Studies, Los Alamos National Laboratory,
Los Alamos, New Mexico 87545}

\date{\today}

\maketitle

\begin{abstract}
A cluster update (the ``operator-loop'') is developed within the framework 
of a numerically exact quantum Monte Carlo method based on the power 
series expansion of $\rm{exp}(-\beta H)$ (stochastic series expansion). 
The method is generally applicable to a wide class of lattice Hamiltonians
for which the expansion is positive definite. For some important models 
the operator-loop algorithm is more efficient than loop updates previously 
developed for ``worldline'' simulations. The method is here tested on 
a two-dimensional anisotropic Heisenberg antiferromagnet in a magnetic field.

\end{abstract}

\vskip2mm]

The path integral formulation of quantum statistical mechanics 
is a useful starting point for numerical studies of interacting many-body 
systems in cases where positive-definiteness can be assured. Monte 
Carlo algorithms based on the ``Trotter decomposition'' 
\cite{suzuki,worldline} in discrete imaginary time, commonly
referred to as ``worldline'' methods, have been used extensively for 
studies of quantum spins and bosons, as well as fermions in 
one dimension (in higher dimensions the fermion path integral is not 
positive definite).\cite{reviews} Recently, two important technical 
developments have lead to significantly more efficient simulation algorithms.
A generalization \cite{evertz} of cluster updates used in classical 
Monte Carlo simulations \cite{swendsen} can reduce the 
autocorrelation times of some simulations by orders of magnitude, 
\cite{wiese,kawashima} thereby enabling studies of models in parameter 
regimes where  standard local updating schemes do not efficiently explore 
the configuration space. Algorithms have also been constructed that work 
directly in the imaginary time continuum, \cite{sse1,prokofev,beard,irsse}
thus producing results free of systematic errors without the extrapolations
to zero discretization which are required in order to obtain numerically exact 
results using the Trotter decomposition. 

There are, however, still unresolved issues for these improved 
algorithms. For some important models the loop schemes do not take 
into account all interactions in the system, and hence an {\it a posteriori}
acceptance probability has to be assigned after the loop-clusters have been 
constructed.\cite{kawashima,troyer} This can seriously affect the
efficiency of simulations. Some loop algorithms also break down due to 
``freezing'',\cite{evertz,kohno} when the probability is high for 
a single cluster to encompass the whole system. It is also often 
a highly non-trivial task to construct an algorithm for a new Hamiltonian 
--- it would clearly be desirable to have a simple recipe for an
arbitrary model.

In this Communication, a general loop-type updating scheme is constructed 
within the ``stochastic series expansion'' (SSE) \cite{sse1,sse2} framework.
This approach to quantum simulations is based on sampling the diagonal
matrix elements of the power series expansion of ${\rm exp}(-\beta H)$ 
[where $H$ is the Hamiltonian and $\beta$ the inverse temperature] and 
is related to a less general method proposed by Handscomb.\cite{handscomb} 
The SSE scheme is as general in applicability as the worldline 
method, and like the continuous time variant, it is numerically exact 
(there is also a strong relationship between the two methods \cite{irsse}). 
SSE algorithms have been applied to numerous problems over the past several 
years, but so far only local updating schemes have been used. The 
``operator-loop'' algorithm introduced here has the same favorable
effects on autocorrelation times as the loop updates developed 
within the worldline scheme. In addition, the method overcomes the problems 
discussed above; all interactions are taken into account in the loop 
construction, there does not appear to be any problems related to freezing,
and the algorithm is very easily implemented for a wide range of models. 

For definiteness and sake of simplicity, the operator-loop algorithm
will here be described for simulations of the anisotropic $S=1/2$ 
Heisenberg model in a magnetic field, defined in standard notation 
by the Hamiltonian
\begin{equation}
H  =  J\sum\limits_{\langle i,j\rangle} [\Delta S^z_i S^z_j + 
\hbox{$1 \over 2$}(S^+_iS^-_j + S^-_iS^+_j)] - h\sum\limits_i S^z_i ,
\label{modelh}
\end{equation}
where $\langle i,j\rangle$ denotes a pair of interacting spins on a 
lattice in any number of dimensions. In addition to serving as 
an illustration for a general SSE operator-loop algorithm, 
simulation results for this model will show explicitly that problems 
present with other loop algorithms are avoided. With the standard 
worldline loop algorithms, freezing occurs for $\Delta > 1$.
\cite{evertz,kohno} The loop construction also does not take into 
account a non-zero magnetic 
field $h$,\cite{troyer} hence making simulations of large $h >0$ systems 
problematic. In the present algorithm, $h$ is explicitly taken into 
account in the loop construction and simulation results show that 
$\Delta > 1$ poses no problems.

For the construction of the SSE configuration space the Hamiltonian is
first written as
\begin{equation}
H = -J\sum\limits_{b=1}^M [H_{1,b} - H_{2,b}],
\end{equation}
where $H_{1,b}$ and $H_{2,b}$ are symmetric bond operators corresponding to
an interacting spin pair $\langle i(b),j(b)\rangle$;
\begin{eqnarray}
H_{1,b} & = & C - \Delta S^z_{i(b)}S^z_{j(b)} + 
\hbox{$h\over 2J$} (S^z_{i(b)}+S^z_{j(b)}) \nonumber \\
H_{2,b} & = & \hbox{$1\over 2$}(S^+_{i(b)}S^-_{j(b)} + 
S^-_{i(b)}S^+_{j(b)}).
\label{hbs}
\end{eqnarray}
The constant $C$ only shifts the energy and can be chosen to assure
a positive definite expansion for any non-frustrated lattice. The number 
of spins in the system is denoted by $N$; the number of bonds 
$M=Nd$ for a cubic lattice in $d$ dimensions.

The partition function $Z={\rm Tr}\lbrace {\rm e}^{-\beta H}\rbrace$
is expanded as
\begin{equation}
Z = \sum\limits_\alpha \sum\limits_{n=0}^\infty {(-\beta)^n\over n!}
\langle \alpha | H^n |\alpha \rangle ,
\label{zn}
\end{equation}
in the basis $\lbrace |\alpha \rangle \rbrace = 
\lbrace | S^z_1,S^z_2,\ldots,S^z_N 
\rangle \rbrace$. This expansion converges exponentially for $n \sim N\beta$.
A truncation at $n=L$ of this order is imposed, and a unit operator 
$H_{0,0}=1$ is introduced to rewrite Eq.~(\ref{zn}) as (for a more 
thorough discussion, see Ref.~\onlinecite{sse2})
\begin{equation}
Z = \sum\limits_\alpha \sum\limits_{S_L} {\beta^n (L-n)! \over L!}
\left\langle \alpha \left | \prod_{i=1}^L H_{a_i,b_i} \right | 
\alpha \right \rangle ,
\label{zl}
\end{equation}
where $S_L$ denotes a sequence of operator-indices; 
\begin{equation}
S_L = [a_1,b_1]_1,[a_2,b_2]_2,\ldots ,[a_L,b_L]_L,
\end{equation}
with $a_i \in \lbrace 1,2\rbrace$ and $b_i \in \lbrace 0,\ldots,M \rbrace$,
or $[a_i,b_i]=[0,0]$, and $n$ denotes the number of non-$[0,0]$
elements in $S_L$. In principle, each term in (\ref{zl}) should be 
multiplied by a factor $(-1)^{n_2}$, where $n_2$ is the total number of 
$[2,b]$ elements in $S_L$. However, for a non-frustrated lattice this 
number must always be even for the matrix element to be non-zero. Choosing 
$C$ in (\ref{hbs}) such that all matrix elements of $H_{1,b}$ are positive, 
the expansion is then positive definite. A Monte Carlo procedure can 
therefore be used to sample the terms $(\alpha,S_L)$ according to their 
relative weights. Previously,\cite{sse1,sse2} sampling schemes were 
devised based on (i) local substitutions of single diagonal operators, $[0,0]_p
\leftrightarrow [1,b]_p$, and (ii) pairs of diagonal and off-diagonal 
operators $[1,b]_{p_1} [1,b]_{p_2} \leftrightarrow [2,b]_{p_1} [2,b]_{p_2}$.
The diagonal update (i) will also be used here. The new operator-loop update 
involves any number of diagonal and off-diagonal operators and is much 
more efficient than the simple pair update (ii).

It is convenient to introduce the notation $|\alpha (p)\rangle$ for states 
obtained by acting on $|\alpha \rangle$ in Eq.~(\ref{zl}) with the first $p$
elements of the operator string,
\begin{equation}
|\alpha (p)\rangle \sim \prod\limits_{i=1}^p H_{a_i,b_i} |\alpha\rangle ,
\label{propagated}
\end{equation}
and to define states $|\alpha_b(p)\rangle = 
|S^z_{i(b)}(p),S^z_{j(b)}(p)\rangle$ on the bonds. For a contributing
term, $|\alpha (L) \rangle = |\alpha (0) \rangle = |\alpha \rangle$.

The simulation starts with some random state $|\alpha \rangle$ and an
operator string $[0,0]_1,\ldots,[0,0]_L$ containing only unit operators.
The cut-off $L$ is chosen arbitrarily and adjusted during the equilibration
phase of the simulation so that it will always be larger than the highest $n$
reached (hence leading to no detectable truncation error). 
The diagonal update $[0,0]_p \leftrightarrow [1,b]_p$ is carried out 
sequentially at each position $p=1,\ldots ,L$ for which $[a_p,b_p]=[0,0]$ 
or $[1,b]$. When accepted, such an update changes the 
expansion power $n$ by $\pm 1$. Acceptance probabilities that satisfy
detailed balance are obtained using Eq.~(\ref{zl}) and the fact that
there are $M$ random choices for $b$ in the $\to$ direction;\cite{sse2}
\begin{eqnarray}
P([0,0]_p \to [1,b]_p) & = & 
{M\beta \langle \alpha_b(p)| H_{1,b} | \alpha_b(p) \rangle \over L-n }
\nonumber \\
P([1,b]_p \to [0,0]_p) & = &
{L-n+1 \over M\beta \langle \alpha_b(p)| H_{1,b} | \alpha_b(p) \rangle },
\label{diap}
\end{eqnarray}
where a number larger than $1$ should be interpreted as probability one.
The state $|\alpha (0)\rangle $ is stored at the beginning of an
updating cycle. Each time an off-diagonal operator $[2,b]_p$ is
encountered, the corresponding spins are flipped so that the states
in Eqs.~(\ref{diap}) will be available when needed. 

The second, new type of update is carried out with $n$ fixed. It is then
convenient to disregard the $[0,0]$ unit operator elements in $S_L$
and instead work with sequences $S_n$ containing only the Hamiltonian
operators $[1,b]$ and $[2,b]$. The propagation index $p$ will in the
following refer to this reduced sequence. Further, full bond
operators including both the diagonal and off-diagonal terms
are defined; $H_b = H_{1,b}+H_{2,b}$. The matrix element in 
(\ref{zl}) can then be written as
\begin{equation}
M(\alpha,S_n) =
\prod\limits_{p=1}^n 
\langle \alpha_{b_p} (p) | H_{b_p} | \alpha_{b_p}(p-1) \rangle .
\label{mproduct}
\end{equation}
The non-zero matrix elements are
\begin{eqnarray}
\langle \downarrow ,\downarrow | H_b | \downarrow ,\downarrow \rangle &=&
C - \Delta /4 - h/(2J), \nonumber \\
\langle \uparrow ,\uparrow | H_b | \uparrow ,\uparrow \rangle &=&
C - \Delta /4 + h/(2J),  \nonumber  \\
\langle \downarrow ,\uparrow | H_b | \downarrow ,\uparrow \rangle &=&
\langle \uparrow ,\downarrow | H_b | \uparrow ,\downarrow \rangle =
C + \Delta /4, \label{matrelem}  \\
\langle \uparrow ,\downarrow | H_b | \downarrow ,\uparrow \rangle &  = &
\langle \downarrow ,\uparrow | H_b | \uparrow ,\downarrow \rangle = 1/2.
\nonumber
\end{eqnarray}
C should be chosen such that all diagonal matrix elements are larger
than (or equal to) zero. $M(\alpha,S_n)$ can be graphically represented 
as a set of $n$ vertices connected to the propagated spins, as shown 
in Figure 1(a) for a system with 4 spins. Two spin states ``enter'' 
each vertex, and ``exit'' in either the same states or flipped 
(the direction of the propagation is clearly irrelevant). The allowed
types of vertices, corresponding to the non-zero matrix elements
(\ref{matrelem}), are shown in Fig.~1(b). Fig.~1(a) displays all the 
full spin states at each ``event'', but clearly there is much redundant 
information in this picture. The spin states at the four ``legs'' of 
the $n$ vertices completely specify the full spin configuration (except
for spins that happen not to be connected to any vertex). In order to 
carry out the operator-loop update, a linked list of the vertices with 
their four spin states is constructed using the current state 
$|\alpha \rangle$ and the index sequence $S_L$. The list is doubly 
linked, so that it is possible to move in either direction from any 
leg of a given vertex to the leg of the next or previous vertex 
connected to the same spin. 

\begin{figure}
\centering
\epsfxsize=4cm
\leavevmode
\epsffile{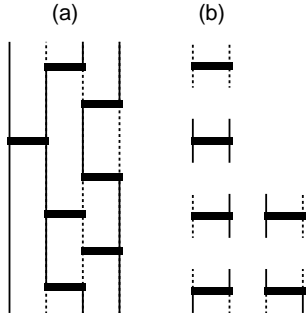}
\vskip1mm
\caption{(a) Representation of a matrix element product $M(\alpha,S_n)$,
Eq.~(\protect{\ref{mproduct}}), with $n=7$, for a 4-spin system. The
vertical solid and dashed lines indicate the spin states acted on by
the operators $H_b$, which are represented by the horizontal bars. 
(b) shows the allowed vertices, which are associated with the non-zero
matrix elements (\protect{\ref{matrelem}}).}
\label{fig1}
\end{figure}

\begin{figure}
\centering
\epsfxsize=6.5cm
\leavevmode
\epsffile{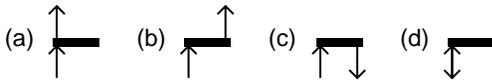}
\vskip1mm
\caption{The four paths on a vertex in the case of the entrance
point being the low-left leg. The entrance and exit legs are indicated
by the arrows. The spins at these legs are flipped in the process; the
states at the other legs remain unchanged.}
\label{fig2}
\end{figure}

The principles of the operator-loop update are now quite simple to state: 
One of the $n$ vertices is first chosen at random, and one of its four 
legs is randomly selected as the entrance point. One of its legs is 
then chosen as the exit point from the vertex, according to probabilities
to be specified below. The four possible vertex paths, in the 
case of the entrance being the low-left leg, are illustrated in Figure 2.
The spins at both the entrance and exit legs are flipped. [Note that the 
entrance and the exit can be the same leg, Fig.~2(d), in which case 
the net effect is no spin flip; only a reversal of direction of movement in 
the list.] The chosen exit leg points to a leg of another vertex in the 
linked list, the spin at which is also flipped. From this vertex, an exit 
leg is again chosen, which points to another vertex, e.t.c.. After some 
varying number of steps, the exit of the last visited vertex will point 
to the original entrance point of the update. The loop then 
closes and the result has been to flip all the spins along the random 
path followed in the process. Since the operator list is a periodic
structure (because $|\alpha (n) \rangle = |\alpha (0)\rangle$), any
state $|\alpha (p)\rangle$ can be affected in the update, and the sum over
states $|\alpha \rangle$ in Eq.~(\ref{zl}) is therefore, implicitly, 
also sampled in the process.

The probabilities for the four different choices of exits from a given 
visited vertex are simply proportional to the matrix elements (\ref{matrelem})
corresponding to the resulting vertices, i.e., those where the spins
at the entrance and exit legs have been flipped. It is intuitively clear that 
this operator-loop procedure satisfies detailed balance and, in combination
with the diagonal single-operator update, is ergodic in the grand canonical 
ensemble (fluctuating total $z$-component of the magnetization), including 
all winding number sectors. For lack of space, a rigorous proof will 
not be presented here. 

Note that one of the paths (a)-(c) in Fig.~2 will always have zero 
probability, since the Hamiltonian (\ref{modelh}) does not contain
operators $S^+_iS^+_j$ or $S^-_iS^-_j$. These operators could be 
included in a more general model and then all four paths would 
be allowed. The probability of the ``bounce'' process (d) is always 
in principle non-zero. However, in some cases it is possible to exclude
this path. Consider the $XY$ model in zero field, i.e., 
$\Delta=h=0$. If $C=1/2$ is chosen, all the non-zero matrix elements in 
(\ref{matrelem}) equal $1/2$. Detailed balance is then satisfied 
also by only choosing, with equal probabilities, among the two allowed
paths (a)-(c). For the isotropic Heisenberg model, i.e., $\Delta=1$, $h=0$, 
and with $C=1/4$, the bounce can also be neglected. The only allowed path
is then always the ``switch and reverse'' (c) [which corresponds
to a substitution $[1,b] \leftrightarrow [2,b]$ in terms of the operators 
in $S_L$], and hence the loop construction is completely deterministic 
in this important case. 

A full updating cycle consists of the following steps: First the diagonal
single-operator update is carried out at all positions in $S_L$ with
diagonal operators. The linked list of vertices is then constructed and
a number of loop updates are performed. The typical size of a loop 
depends strongly on the model parameters. The number of loops to be 
constructed in each cycle is therefore chosen such that on average 
a total of $\sim \langle n\rangle$ vertices are visited. The updated vertices 
are finally mapped onto the corresponding operator-indices $[a,b]$
and written into $S_L$.

To demonstrate the efficiency of the new algorithm, results are 
next presented for two different cases where previous loop algorithms 
have encountered difficulties:\cite{troyer,kohno} The anisotropic model 
in zero field and the isotropic case with a field. The estimators for 
various observables of interest have been discussed in detail in 
Ref.~\onlinecite{sse2}. The correctness of the simulation code 
was verified by comparing results for a $4 \times 4$ lattice with exact 
results obtained by diagonalizing the Hamiltonian. The results to be 
presented next were obtained using lattices sufficiently large to eliminate 
finite-size effects. For the lowest temperatures considered, $64\times
64$ spins were typically used, and on the order of $2 \times 10^6$
updating cycles were carried out.

\begin{figure}
\centering
\epsfxsize=8.2cm
\leavevmode
\epsffile{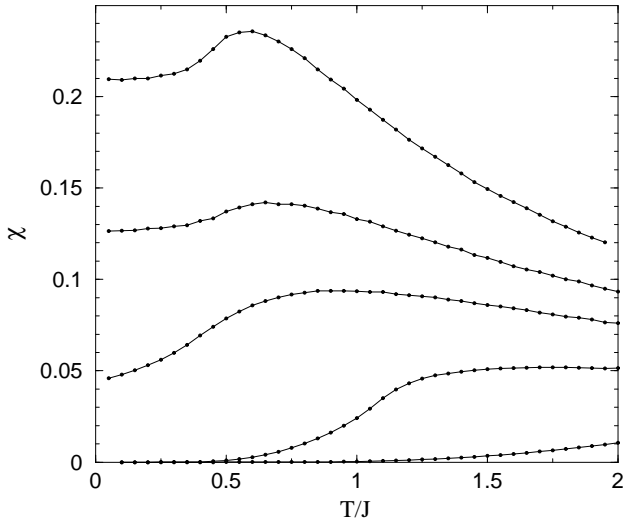}
\vskip1mm
\caption{Magnetic susceptibility vs temperature for the zero-field
Heisenberg model with anisotropy parameter $\Delta=0$, $0.5$, $1$, 
$2$, and $4$ (top to bottom).}
\label{fig3}
\end{figure}

The susceptibility, $\chi = \beta \langle (\sum_i S^z_i)^2 \rangle /N$, 
for the case $h=0$ is shown in Figure \ref{fig3} for several values of 
the anisotropy $\Delta$. Unlike with the standard worldline loop 
algorithm,\cite{evertz,kohno} there are no problems with ``freezing'' 
in simulations for $\Delta > 1$. The exponential decay of $\chi$
to $0$ as $T \to 0$ for $\Delta > 1$ reflects the opening of a gap in 
the spectrum for these systems. For the isotropic case ($\Delta=1$), 
the results are in perfect agreement with previous calculations.\cite{kim} 
For the $XY$ model ($\Delta=0$), a temperature-independent behavior is
seen at low-temperature $(T/J \alt 0.2$), in agreement with a prediction 
of chiral perturbation theory.\cite{chiral} Quantitatively, the $T$-independent
value should be $\chi = \rho_s/c^2$,\cite{chiral} where $\rho_s$ is 
the spin stiffness and $c$ the spin-wave velocity. The result 
$\chi = 0.2095(3)$ obtained here at $T/J=0.05$ is consistent with 
this prediction and recent ground state calculations of 
$\rho_s$ and $c$.\cite{hamer}

\begin{figure}
\centering
\epsfxsize=8.2cm
\leavevmode
\epsffile{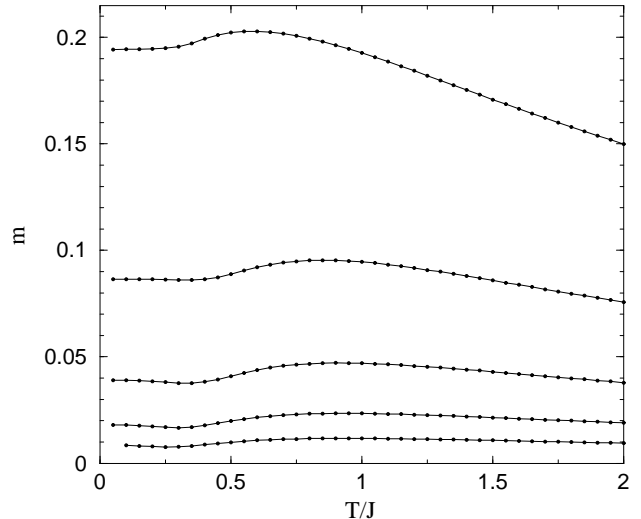}
\vskip1mm
\caption{Magnetization vs temperature for the isotropic Heisenberg
model ($\Delta=1$) in magnetic fields $h/J=2$, $1$, $1/2$, $1/4$, 
and $1/8$ (top to bottom).}
\label{fig4}
\end{figure}

The magnetization per spin, $m=\langle \sum_i S^z_i \rangle /N$, is shown 
for an isotropic interaction and several strengths of the magnetic 
field in Figure \ref{fig4}. For all field-strengths, there is a
maximum in $m$ between $T/J=0.5$ and $1$, reflecting the cross-over between 
high-temperature independent spin behavior and antiferromagnetic correlations
developing at lower $T$ (also seen in the zero-field susceptibility 
in Fig.~\ref{fig3}). Note the shallow minimum at lower temperatures 
for $h/J \le 1$. This reflects the temperature scale at which the 
local, short-range antiferromagnetic correlations are the strongest. 

The operator-loop simulations are very efficient for any strength of
the field, since a $h > 0$ is taken into account in the loop
construction. With other loop algorithms,\cite{evertz,beard}
an {\it a posteriori} acceptance probability has to be assigned
for updates in which the total magnetization changes. This acceptance
probability decreases rapidly with increasing field strength, leading 
to an autocorrelation time which increases exponentially with 
$h/T$.\cite{troyer} Previous simulations \cite{troyer} were therefore 
restricted to $h/T \alt 4$. Fig.~\ref{fig4} shows results up to $h/T=40$, 
and there are no signs of increasing autocorrelation times even for 
much higher values.

To conclude, the operator-loop algorithm introduced here has several 
advantages over other loop methods suggested recently.\cite{evertz,beard}
The most important is that all interactions, including external fields, 
are taken into account in the loop construction, thus eliminating the need 
for {\it a posteriori} acceptance probabilities that restrict the 
applicability of the previous methods.\cite{troyer} Like the continuous-time
version of the worldline algorithm, \cite{prokofev,beard,irsse} the SSE 
method is completely approximation free. The configuration space is, 
however, discrete, and the only floating point operation required in
the simulation is the generation of uniformly distributed random numbers. 
In the continuous-time worldline algorithms,\cite{prokofev,beard,irsse} 
on the other hand, high-precision values of imaginary times have to 
be manipulated. One can therefore expect that the operator-loop algorithm 
is faster in many cases, in particular for the uniform Heisenberg model,
where the loop construction is deterministic. It is also interesting 
to note that certain expectation values have simpler estimators in the 
SSE framework than for worldline methods.\cite{irsse}

The method has here only been demonstrated for the anisotropic Heisenberg 
in a magnetic field. Generalizations to other models with two-body 
interactions are almost trivial, however. The vertices depicted 
in Fig.~1 only involve other degrees of freedom at
the ``legs''. For example, for Hubbard-type models the legs can have
charge $c=1$ and spin $s=\pm {1\over 2}$, or $s=0$ and $c=0,2$. The 
vertex paths in Fig.~2 then involve changing these quantum numbers
by some values $(\delta c,\delta s)$ at the entrance leg, and changing 
them by $(\delta c,\delta s)$ at an exit leg in the same direction [paths (a) 
and (b) in Fig.~2] or $(-\delta c,-\delta s)$ at an exit in the reverse 
direction [paths (c) and (d)]. Implementation for a new model thus
essentially involves specifying all allowed vertices, i.e., all non-zero
matrix elements of type (\ref{matrelem}). 

This work is supported by the NSF under Grant No.~DMR-9712765. Part of
the research was carried out during visits at the {\it Institut Romand 
de Recherche Num\'{e}rique en Physique des Materiaux} (IRRMA), Lausanne, 
Switzerland, and the School of Physics, the University of New South Wales,
Sydney, Australia. I thank these institutions for their hospitality 
and financial support.

\end{document}